\documentclass[twocolumn,showpacs,preprintnumbers,amsmath,amssymb]{revtex4}

\usepackage{graphicx}
\usepackage{dcolumn}
\usepackage{bm}

\begin{document}


\title{Magic number 7 $\pm$ 2 in networks of threshold dynamics}

\author{Shuji Ishihara}
 \email{shuji@complex.c.u-tokyo.ac.jp}
\affiliation{
  Department of Pure and Applied Sciences, University
  of Tokyo, \\
  Komaba, Meguro-ku, Tokyo 153-8902, Japan \\
}
\author{Kunihiko Kaneko}
\affiliation{
  Department of Pure and Applied Sciences, University
  of Tokyo, \\
  Komaba, Meguro-ku, Tokyo 153-8902, Japan \\
}

\date{\today}

\begin{abstract}
Information processing by random feed-forward networks consisting of
units with sigmoidal input-output response is studied by focusing on
the dependence of its outputs on the number of parallel paths M.  
It is found that
the system leads to a combination of on/off outputs when $M \lesssim 7$,
while for $M \gtrsim 7$, chaotic dynamics arises, resulting in  a 
continuous distribution of outputs.  This universality of the critical
number $M \sim 7$ is explained by combinatorial explosion, i.e.,
dominance of factorial over exponential increase.  Relevance of the
result to the psychological magic number $7 \pm 2$ is briefly discussed.
\end{abstract}


\maketitle

Information processing (IP) in biological systems is often carried out by
elements that show threshold-type (sigmoid) input-output (IO)
behaviors.  For example, the expression of a gene is determined by
an on/off-switch
for its transcription factors.  Another example is the signal transduction 
system in a cell where the enzymatic reaction often displays a sigmoid
output when responding to an external stimulus like the abundance of an 
input chemical\cite{signal}. 
One of the most well-studied examples of
sigmoid IO relationships can be found in neural networks, where the output of
each neuron depends on inputs from other neurons or a receptive field
\cite{Rosensenblatt}.

In these biological networks, the connections among elements are often
entangled.  Cross-talk in signal transduction has recently been observed
for several systems\cite{signal}, while enzymatic reactions are generally
entangled as well.  The connections in neural networks are known to be
complex.  Besides complexity, biological networks can also display
cascade type structures leading from the external input to the final output.
In signal transduction such cascades are encountered, while
 layered networks have been discussed as an idealization of
biological neural networks.  Hence, the study of entangled
layered networks is generally important.  The information processing in
such systems is discussed by judging whether distinct attracting points
(or sets) are reached through the dynamics in successive layers,
depending on the input.  In a layered network system, e.g., the 
attracting set is the state of the output layer.

In an entangled network, the more the number of degrees of freedom to be processed
increases, the more mutual interference can occur thus increasing the complexity.  
Consequently, the IP ability of the network depends on the number of
processed degrees of freedoms.   In this paper, we discuss this number dependence
and explore some universal properties of entangled networks with sigmoid units.

In connection with this problem, it is interesting to note that the
term magic number $7 \pm 2$ was coined in psychology \cite{Miller},
where the number of chunks (items) that can be memorized in short term
memory is found to be limited to about $7 \pm 2$.  In
neural networks, this corresponds to the number of inputs
beyond which the output that depends on these inputs no longer clearly
separates.

In order to investigate the question raised above, we adopt a cascade perceptron
as an abstract model of a random sigmoid response\cite{McCulloch-Pitts,Minsky}.
Here we consider feed-forward network dynamics without feed-back loop,
for simplicity.  Each layer $l$  is composed of $M$ elements and all
elements are regulated by the elements in the preceding layer:
\begin{eqnarray}
  x^l_{i} = \tanh \left( \frac{\beta }{\sqrt{M}} \sum_{k}^M \epsilon^l_{ik}
  x^{l-1}_{k} - \theta^l_{i} \right), \label{eq:model}
\end{eqnarray}
where $x^l_{i}$ represents the state of the $i$-th element of the $l$-th
layer. $\theta^l_{i}$ is the threshold value for $x^l_{i}$ to be `excitatory'.
Unless otherwise stated, we set $\theta^l_{i} =0$ 
as this specific choice is not important for the later
discussion. The coupling terms $\epsilon^l_{ik}$ are chosen
randomly from a Gaussian distribution with standard deviation $1.0$.
The parameter $\beta$ normalized by $\sqrt{M}$
determines the steepness of the sigmoid function.  As
$\beta$ approaches $0$, 
Eq.(\ref{eq:model}) approaches a constant function with
$x^l_{i} = 0$ (or $x^l_{i} = -\theta^l_{i}$).  On the
other hand, as $\beta$ increases,  equation (1) approaches a
step functions such that in almost all cases $x^l_{i}$ becomes 
either $-1$ or $1$, and each element effectively has just $2$ states. 
For the medium range of $\beta$,
the IO relationship is smooth, which, as will be shown later, may lead
to complex dynamics.  Note that if all the thresholds
$\theta^l_{i}=0$, the change of sign $x \to -x$ preserves the equations
of the system, so that the solutions for
Eq.(\ref{eq:model}) are symmetric.

For the information processing stages carried out in each layer,
$\vec{x}^l = F_{l}(\vec{x}^{l-1})$ holds where
$\vec{x}^l \equiv \left( x^l_{1},x^l_{2},\cdots,x^l_{M}\right)$ and
$F_l$ is the processing dynamics carried out in the $l$-th layer.  We set
the values of the $0$-th layer as the inputs
to be processed. 
If a succeeding layer is regarded as a next time step,
the present system can be interpreted as a random dynamical system\cite{Ott.PRL}
where the processing corresponds to  temporal evolution.
We take various inputs (corresponding to a {\it set of
inputs}) randomly chosen such that $x^0_{i} \in [-1,1]$ for each
$i$, and compute numerically the evolution of Eq. (\ref{eq:model}).

\begin{figure}[tbhp]
\centering
\includegraphics[width=.45\textwidth]{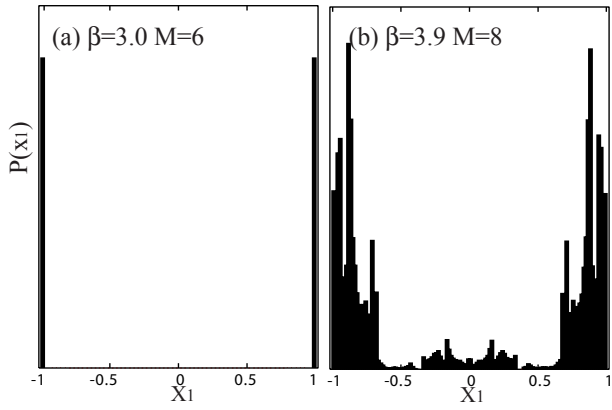}
\caption{ Distribution of $P(x^{l}_1)$ at layer$l=30$, obtained
  over $5.0\times10^5$ inputs with a homogeneous distribution over
  $[-1,1]$. The parameter $\beta=3.0$ and the threshold $\theta$ is set
  to $0$.  The histogram is plotted by using a bin size $2.0/128$.
  (a) for $M=6$ (b) $M=8$.  }
\label{fig:1bodyDist}
\end{figure}

Let us first discuss the qualitative behavior of Eq.(\ref{eq:model}).
For $\theta^l_i=0$, $x^l_i$  the outputs converge to $0$,
irrespectively of the inputs if $\beta<1$, while they approach either
$1$ or $-1$ when $\beta$ is large.  For  middle range values of $\beta$,
outputs may take values between -1 and 1, depending on the input.
In this regime, outputs are often sensitive to changes of the inputs, 
and indeed orbital instability exist in the evolution through the layers.
The degree of this instability depends on the number of parallel paths and
on interference.  For sufficiently small $M (\lesssim 7)$, the IP is
stable, in the sense that the $x^l_{i}$ in the output layer
only assume a few distinct values, depending on the input values. 
On the other hand, if
$M$ is large ($\gtrsim7$), such convergence is not common. 
We have computed a histogram of the output values
$P(x^{l}_1)$ sampled over $5.0\times 10^5$ 
randomly chosen inputs.  As shown in Fig.\ref{fig:1bodyDist}, there are
clearly two peaks at $x = \pm 1$ for $M=6$ and $P(x) \approx 0$ for
$x\neq 1$, while for $M=8$, the distribution is broad.

\begin{figure}[tbhp]
\centering
 \includegraphics[width=.4\textwidth]{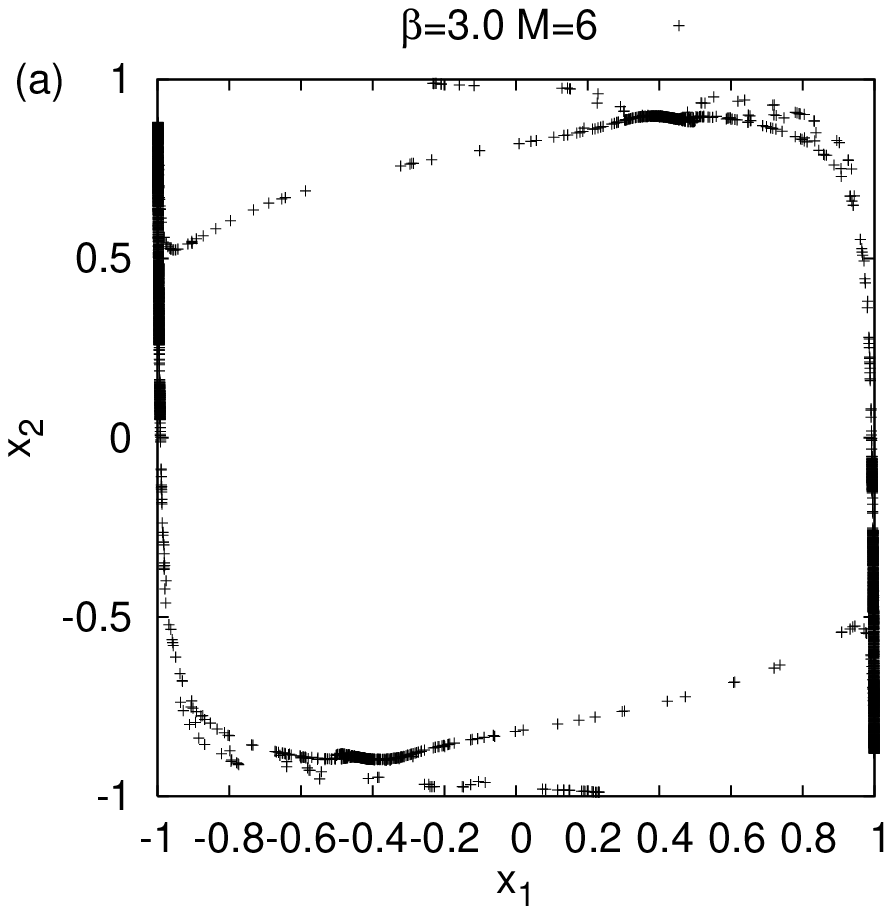}
 \includegraphics[width=.4\textwidth]{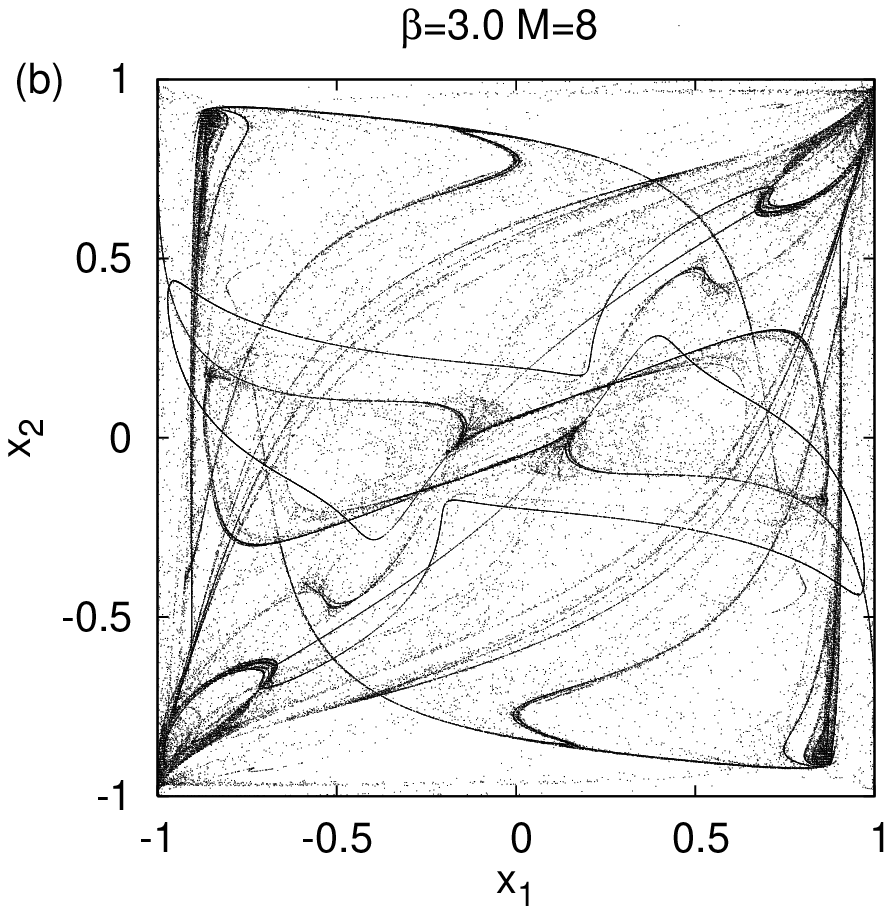}
\caption{ Snapshots of $(x_1^l,x_2^l)$ at layer $l=30$,
  plotted as a projection of $\vec{x}^l$ on to the two dimensional
  plane $(x_1,x_2)$. The data obtained from $10^5$ randomly chosen inputs are
  overlaid with $\beta=3.0$.  (a) $M=6$.  The density of points is
  quite high around $(\pm1,\pm1)$.  (b) $M=8$.  Multiple folding
  and stretching processes are detected, while the points are
  scattered over the plane.}
\label{fig:snapshotL=30}
\end{figure}

The scattering in the values $x^l_{i}$ of the attractor for the latter case is due
to chaotic dynamics in Eq.(\ref{eq:model}), where stretching and
folding in phase space appear \cite{note1}.
Fig.\ref{fig:snapshotL=30}(a,b) show values of $x^l$ projected onto the
$(x^l_1,x^l_2)$ plane.  In the plot we take $l=30$ and $10^5$ inputs
given at $l=0$.  For $M=6$(a) each $x^l_i$ at $l=30$ is localized
within a small volume of the total phase space.  On the other hand, for
$M=8$(b), one can see folding and stretching, and the scattering of
points throughout phase space.  With these chaotic dynamics, tiny
differences in the input values are amplified making clear
separation of inputs impossible.

\begin{figure}[tbhp]
\centering
\includegraphics[width=.45\textwidth]{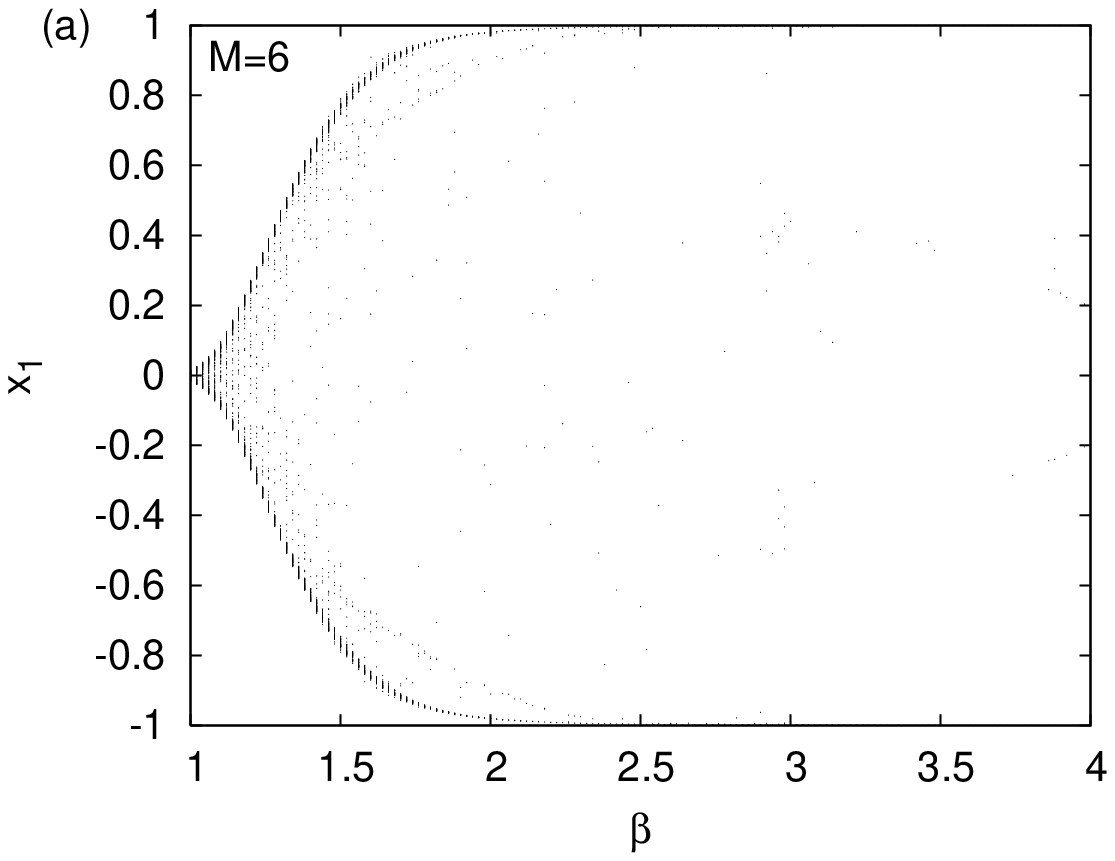}
\includegraphics[width=.45\textwidth]{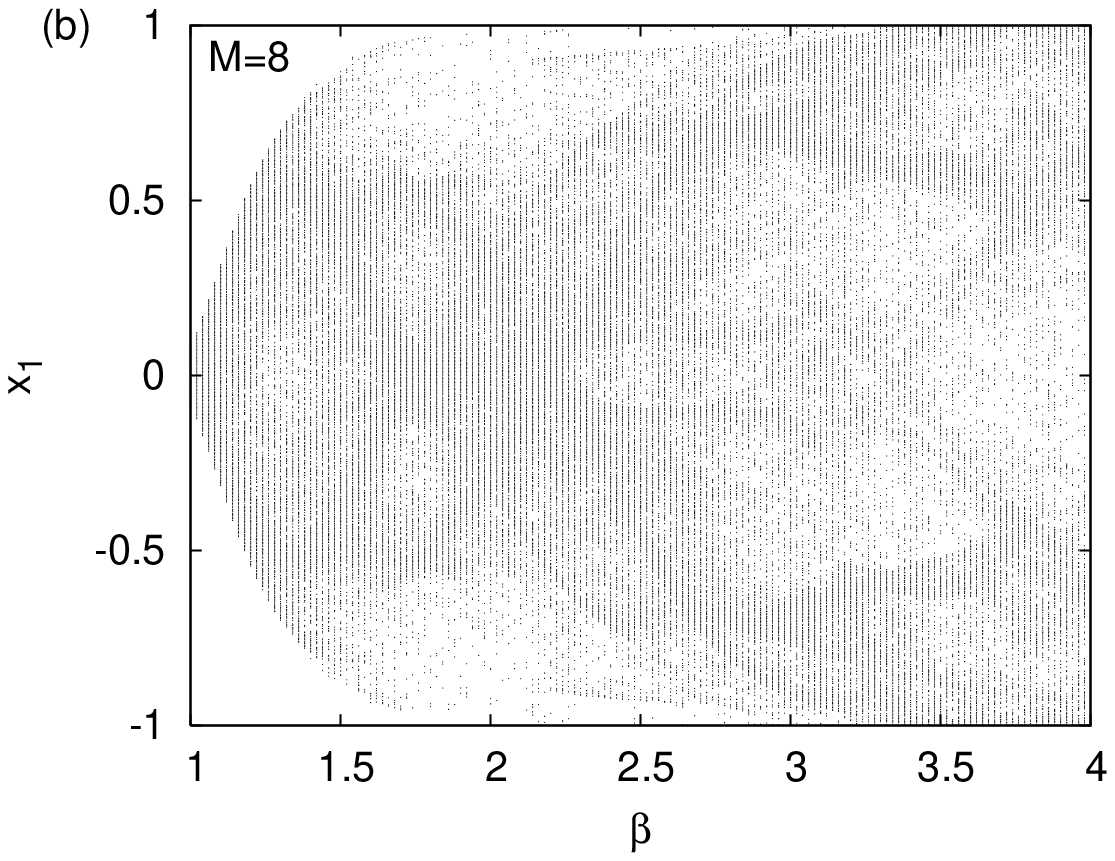}
\caption{ Bifurcation diagram of $x^{l=30}_1$ against $\beta$.
  $x^l_1$ at layer $l=30$ is plotted over $800$ different inputs
  for each $\beta$. For $\beta (\lesssim 1.0)$, $x^l$ converge to
  $x_1=0$. (a) $M=6$. For most inputs, $x^{l=30}_1$ converges to $\pm
  1$ if $\beta$ is sufficiently larger than $1$.  (b) for $M=8$,
  $x^{l=30}_1$ is scattered over $[-1,1]$.  }
\label{fig:bif}
\end{figure}

The above simulations are carried out with $\beta=3.0$, but this
stretching and folding process is observed as long as $\beta(>1)$.
To obtain insight into the dependence on $\beta$, values of $x^{l=30}_1$ for 800
inputs are plotted as a function of $\beta$ in Fig.\ref{fig:bif}.  For $M=6$,
they converge to a few points for a large portion of
$\beta$, while for (b) $M=8$ they do not.

These numerical results suggest that a critical number of
parallel paths $M_c$ exists, beyond which chaotic dynamics is inevitable, and that
$M_C$ is around $7$ for a wide range of $\beta$. We confirm this critical number
by computing several characteristic quantities for the model Eq.(\ref{eq:model}).

\begin{figure}[tbhp]
\centering
\includegraphics[width=.45\textwidth]{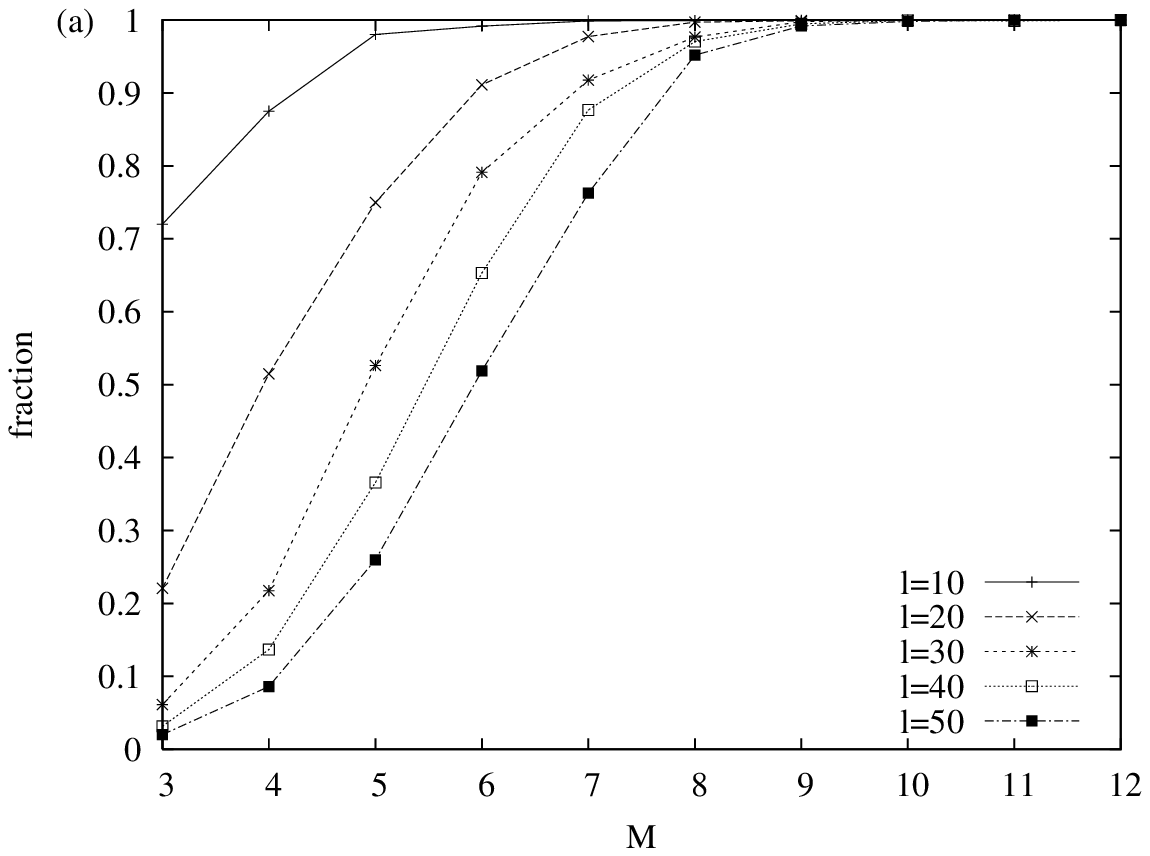}
\includegraphics[width=.45\textwidth]{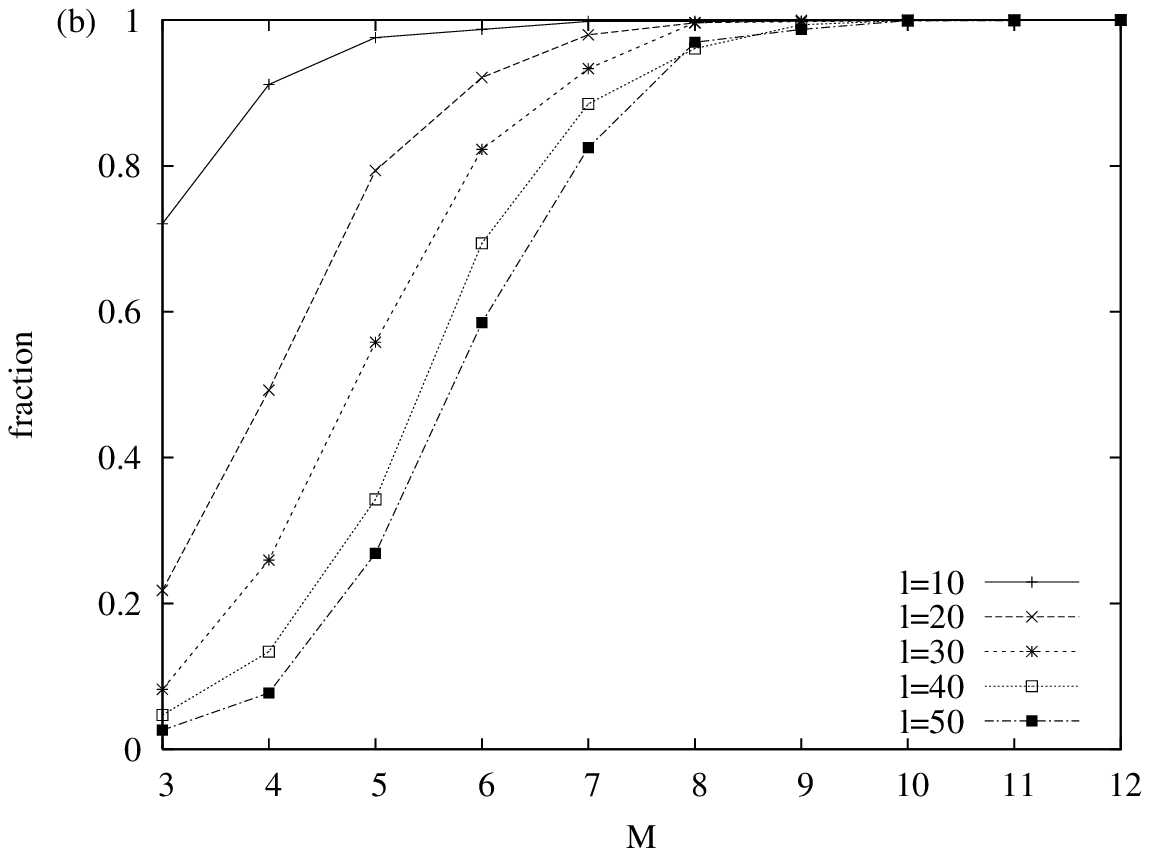}
\caption{ The average fraction of the value $x$ such that
  $P(x^l_1=x)>0$, plotted as a function of $M$.  The histogram is
  computed for $10^5$ inputs over $200$ networks (i.e., with different
  choices of $\epsilon^l_{ik}$), with a bin size $2.0/128$.  The
  fraction at layer $l=10,20,30,40,50$ is plotted.  (a) with
  threshold $\theta^l_i=0$ and (b) with distributed threshold
  $\theta_i \in [0,0.5]$. The parameter $\beta$ is fixed at $3.0$.  }
\label{fig:N-ratio}
\end{figure}

First, we plot the fraction of bins for which $P(x^{l}_1)$ is
not zero in each layer in Fig.\ref{fig:N-ratio}(a). Here
$P(x^{l}_1)$ is computed over $5.0\times 10^5$ inputs, by taking a bin
size of $2.0/128$.  For $M \le 7$ the fraction becomes smaller from
layer to layer, while for $M \ge 8$ the fraction is almost one and does not
decrease much for successive layers.  For $M \le 7$, the output points are well
separated by the sigmoid function, while they are scattered over the
whole range of values $[-1,1]$ for $M \ge 8$.  The data for
Fig.\ref{fig:N-ratio}(a) are obtained for a fixed threshold
$\theta_i=0$, but the conclusion does not change even when the thresholds are
distributed, as shown in Fig.\ref{fig:N-ratio}(b) where
$\theta_i \in [0,0.5]$.  These behaviors are also invariant against
 changes in $\beta$, as long as it is sufficiently larger than 1 but
  not that large for the $\tanh$ function to effectively become a
step function.  Hence the critical number $M_c \approx 7$ is rather
general, without dependencies on the details of the model.

\begin{figure}[tbhp]
\centering
\includegraphics[width=.45\textwidth]{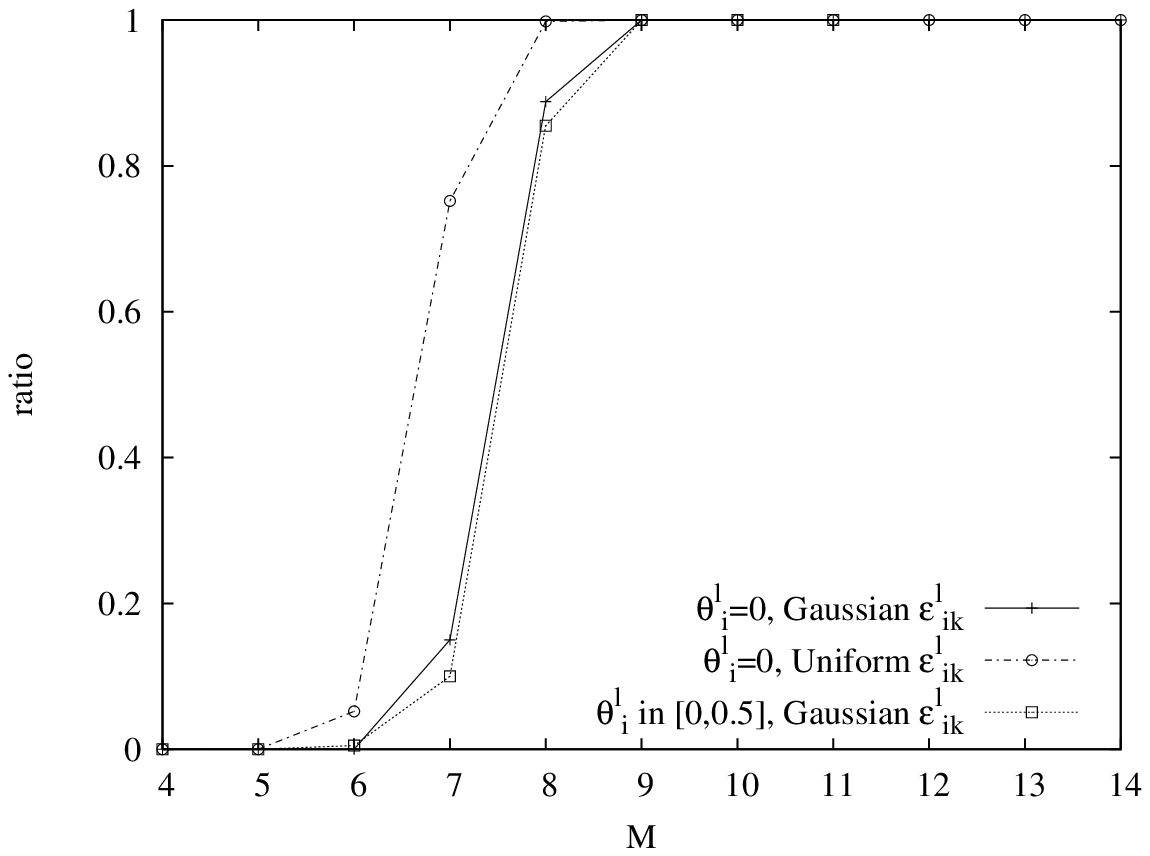}
\caption{ 
  The ratio of networks which shows `chaotic' behavior for
  some values of $\beta$, plotted as a function of $M$.  $500$
  networks are chosen for $\epsilon^l_{ik}$.  The Lyapunov exponent is
  computed by using $J^l$ over 500 steps (layers).  The
  behavior is judged numerically as "chaotic" if the exponent is larger than $0.0$.
  The  ratio is computed for the following three cases; the threshold $\theta^l_i=0$
  with a Gaussian distribution of $\epsilon^l_{ik}$~(cross); distributed
  thresholds $\theta_{ij} \in [0,0.5]$ with a Gaussian distribution of
  $\epsilon^l_{ik}$ ~(box);and $\theta^l_i=0$ with  a uniform
  distribution for $\epsilon^l_{ik} \in [-1,1]$~(circle).  }
\label{fig:M-ratio}
\end{figure}

Second,  we have computed the degree of orbital
instability in the chaotic dynamics, i.e., the sensitivity on
input values.  By regarding a layer as a time step in a dynamical
system, the sensitivity is computed by the Lyapunov exponent of the
random map Eq.(\ref{eq:model}) as follows\cite{stability}:
\begin{eqnarray}
\lambda_{max}= \max_{\vec{x},|\vec{x}|=1}\frac{1}{l} \ln |J^{l}
J^{l-1} \cdots J^{1}\vec{x}| \label{eq:Lyapunov}
\end{eqnarray}
where $J^l$ is the Jacobian matrix of Eq.(\ref{eq:model}),
$J^l_{ik} \equiv \frac{\partial x^l_{i}}{ \partial x^{l-1}_{k}}$,
so that $\delta x^l_{j} = \sum_k J^{l}_{ik}(x^{l-1})~\delta x^{l-1}_{k}$.
The fraction of the network having positive exponents $\lambda_{max}$
is plotted in Fig.\ref{fig:M-ratio} for the following three cases:
$\theta^l_i=0$ with a Gaussian distribution for $\epsilon^l_{ik}$,
$\theta^l_i=0$ with a uniform distribution for $\epsilon^l_{ik} \in
[-1,1]$, and distributed thresholds $\theta_{ij} \in [0,0.5]$ with
a Gaussian distribution for $\epsilon^l_{ik}$.  For all of the three
cases, the fraction of networks with chaotic behavior
drastically increases around $M=7$.

Loss of separability of inputs around the number 7 due to chaotic
dynamics is not limited to the model investigated above.  We have also
investigated some other models consisting of units with threshold
dynamics (of Michaelis-Menten's form for enzymatic reactions), that are
randomly connected in a cascade\cite{IshiharaKaneko}.  The same
behavior with the same critical number 7 is obtained. On the other
hand, it is also interesting to note that Milnor attractors that
collide with their basin boundary are dominant for globally coupled
dynamical systems with more than $7\pm2$ degrees of
freedom\cite{KK-Milnor,KK-magic7}.

Then, why is the critical number $7$ (or $7\pm 2$) so universal?  In
\cite{KK-magic7}, one of the authors(KK) discussed the possibility
that the combinatorial explosion of the basin boundaries due to
chaotic dynamics is relevant to this critical number (i.e., the faster
increase of $N!$ over $2^N$).
This combinatorial argument can be extended to the
present problem.

We do so by considering the origin of the folding process. In order to
see the effect of
entanglement, we study the input-output relationship of
$x^0_1\rightarrow x^2_1$ of a two-layer system\cite{note2} by
fixing the inputs of $x^0_2,\cdots,x^0_{M}$.  Then output $x^2_1$ is
given as a function of $x^0_1$;~$x^2_1=\tanh(\sum_j
\sigma_j(x^0_1-\nu_j))$ where $\sigma_j(u)= \tilde{\beta}
\epsilon^{2}_{1j} \tanh(\tilde{\beta} \epsilon^1_{j1} u)$ and $\nu_j =
(\epsilon^l_{j1})^{-1} \sum_{k=2}^{M} \epsilon^1_{jk} x^0_k$.  
Here it is assumed
that $\tilde{\beta} \equiv \beta/\sqrt{M} $ is large, and that
$tanh(x)$ is close to a step function.  Note that there are $N$ paths
via the middle layer elements where $x^1_j$ switches between the
values -1 and 1 as $x^0_1$ crosses the `threshold' $\nu_j$.  One
can then renumber the index $j=1,\cdots,M$ such that
$-1<\nu_1<\nu_2<\cdots<\nu_{M}<1$.  With this ordering, if $\sigma_j$
is positive and $\sigma_{j+1}$ is negative,  the one-dimensional
mapping $x^0_1 \rightarrow x^2_1$ has a single hump at
$\nu_j<x<\nu_{j+1}$ implying a folding process as in the logistic
map.  Then, if the sign of $\sigma_j$ alternates for successive $j$, 
the above function
switches between $-1$ and $1$ at $\nu_j$ $M$ times as $x^0_1$ is
increased;~ The one-dimensional mapping from the input  $x^0_1$ to the
output $x^2_1$ is thus subject to this folding process everywhere in $-1<x<1$.  Since
$\epsilon^2_{1j}$ can take positive or negative values with equal
probability, the probability to have full folding decreases proportionally to
$2^{-M}$.

The estimate given so far is for fixed inputs of
$x_2,x_3,\cdots,x_{M}$.  By changing these input values, the ordering
of $\nu_j$ changes accordingly (for the original index without
reordering) and hence there are in total $(M-1)!$ possible orderings.
Therefore, roughly speaking, the input-output relationship has full up-down
switches for some input values $x_2,x_3,\cdots,x_{M}$, when
$(M-1)!2^{-M}$ exceeds $1$.  In this case, at every layer, for any
element, the folding occurs fully for some inputs, and the folding
process covers most of phase space.  Even though this argument is quite
rough, it is still possible to presume that when $(M-1)!$ exceeds the
order of $2^M$, the chaotic dynamics replaces the separation by the
threshold function.  Note that this factorial surpasses $2^M$ at
$M=6$ coinciding with the magic number $7 \pm 2$.  
This could be the reason why at the magic number $7\pm 2$, the
separation of states collapses and chaotic dynamics takes over.

In the present paper, we have shown that the interference between inputs
drastically increases around $M \sim 7$ within the general setup of
neural networks.  The argument of magic number $7 \pm 2$ presented 
here is only based
on combinatorial arguments, and does not strongly depend on
the choice of parameters.  Hence it is naturally expected that our explanation
works for a wide class of entangled cascade networks with sigmoid
units. Considering also the generality of the mechanism, it is not particularly
far fetched to infer a correspondence between our result and the original
magic number $7 \pm 2$ in psychology.  Of course, at present
 the underlying neurodynamics associated with the actual psychological
process is still unknown, and hence we do not claim that we have found 
{\sl the} solution to
the magic number 7 problem\cite{TsudaNicolis}.  Nevertheless, the formation of
distinct attracting sets resulting from inputs channeled through layered
networks with sigmoidal elements is common among neural processes,
and hence it is important to mention the connection.

The authors are grateful to K. Fujimoto, T. Shibata and
F. H. Willeboordse for discussion.  This work was supported by a
Grant-in-Aid for Scientific Research from the Ministry of Education,
Science, and Culture of Japan~(11CE2006).

\newpage

\newpage

\newpage

\newpage

\newpage


\begin{thebibliography}{99}
\bibitem{note1}
From a dynamical systems point of view, the above process is
similar to \cite{Ott.PRL} where a change of the parameter value makes
the effective maximum Lyapunov exponent be positive.
\bibitem{stability} 
From this Jacobi matrix, the linear stability of
the state$\vec{x}_l=0$, for the case with $\theta^l_i=0$ is
calculated. This leads to the stability condition for $\beta =
\sqrt{M/2}\exp \left( - \psi (M/2) /2\right)\stackrel{<}{\sim} 1$,
where $\psi(x)$ is a polyGamma function\cite{rand-mat}.
\end{thebibliography}
\end{document}